\documentclass[11pt,onecolumn,draftcls]{IEEEtran}
\usepackage[utf8]{inputenc}
\usepackage{amsmath}
\usepackage{authblk}
\usepackage{acronym}
\usepackage{cite}
\usepackage{url}

\begin{document}

%\begin{acronym}
\newacro{DTW}{Dynamic Time Warping}
\newacro{FFT}{Fast Fourier Transform}
\newacro{HMM}{Hidden Markov Model}
\newacro{KL}{Kullback-Leibler}
\newacro{LSA}{Latent Semantic Analysis}
\newacro{MFCC}{Mel Frequency Cepstral Coefficient}
\newacro{MMR}{Maximal Marginal Relevance}
\newacro{RMS}{Root Mean Square}
\newacro{SVD}{Singular Value Decomposition}
\newacro{SVM}{Support Vector Machine}
\newacro{TF-IDF}{Term Frequency - Inverse Document Frequency}
%\end{acronym}

\title{On the Application of Generic Summarization Algorithms to Music}
\author{Francisco Raposo,
				Ricardo Ribeiro,
				David Martins de Matos,~\IEEEmembership{Member,~IEEE}
\thanks{Francisco Raposo is with Instituto Superior Técnico, Universidade de Lisboa, Av. Rovisco Pais, 1049-001 Lisboa, Portugal}
\thanks{Ricardo Ribeiro is with Instituto Universitário de Lisboa (ISCTE-IUL), Av. das Forças Armadas, 1649-026 Lisboa, Portugal}
\thanks{David Martins de Matos is with Instituto Superior Técnico, Universidade de Lisboa, Av. Rovisco Pais, 1049-001 Lisboa, Portugal}
\thanks{Ricardo Ribeiro and David Martins de Matos are with L2F - INESC ID Lisboa, Rua Alves Redol, 9, 1000-029 Lisboa, Portugal}
\thanks{This work was supported by national funds through FCT -- Fundação para a Ciência e a Tecnologia, under project PEst-OE/EEI/LA0021/2013.}}
\markboth{Submitted to IEEE Signal Processing Letters, VOL. X, No. X}{Submitted to IEEE Signal Processing Letters, VOL. X, No. X}
\maketitle

\begin{abstract}
Several generic summarization algorithms were developed in the past and
successfully applied in fields such as text and speech summarization. In this
paper, we review and apply these algorithms to music. To evaluate this
summarization's performance, we adopt an extrinsic approach: we compare a Fado
Genre Classifier's performance using truncated contiguous clips against the
summaries extracted with those algorithms on 2 different datasets. We show that
\ac{MMR}, LexRank and \ac{LSA} all improve classification performance in both
datasets used for testing.
\end{abstract}

\section{Introduction}
Several algorithms to summarize music have been published
\cite{Chai,Cooper2003,Peeters2002,Peeters2003,Chu2000,Cooper2002,Glaczynski2011,Bartsch2005},
mainly for popular music songs whose structure is repetitive enough. However,
those algorithms were devised with the goal of producing a thumbnail of a song
as its summary, the same way an image's thumbnail is that image's summary.
Therefore, the goal is to output a shorter version of the original song so that
people can quickly get the gist of the whole piece without listening to all of
it. These algorithms usually extract continuous segments because of their human
consumption-oriented purpose.

Generic summarization algorithms have also been developed for and are usually
applied in text summarization. Their application, in music, to extract a
thumbnail is not ideal, because a ``good'' thumbnail entails requirements such
as coherence and clarity. These summaries are composed of small segments from
different parts of the song which makes them unsuitable for human enjoyment and
thus may help evade copyright issues. Nevertheless, most of these algorithms
produce summaries that are both concise and diverse.

We review several summarization algorithms, in order to summarize music for
automatic, instead of human, consumption. The idea is that a summary clip
contains more relevant and less redundant information and, thus, may improve the
performance of certain tasks that rely on processing just a portion of the whole
audio signal. We evaluate the summarization's contribution by comparing the
performance of a Portuguese music style Fado Genre Classifier\cite{Girao2014}
using the extracted summaries of the songs against using contiguous clips (truncated from
the beginning, middle and end of the song). We summarize music using \ac{MMR},
LexRank, \ac{LSA} and also with a method for music summarization called Average
Similarity for comparison purposes. We present results on 2 datasets showing
that \ac{MMR}, LexRank and \ac{LSA} improve classification performance under
certain parameter combinations.

Section \ref{sec:related-work} reviews related work on summarization.
Specifically, the following algorithms are reviewed: Average Similarity in
section \ref{sub:avg-sim}, \ac{MMR} in section \ref{sub:mmr}, LexRank in section
\ref{sub:lexrank} and \ac{LSA} in section \ref{sub:lsa}. Section
\ref{sec:experiments} describes the details of the experiments we performed for
each algorithm and introduces the Fado Classifier. Section \ref{sec:results}
reports and discusses our classification results and section
\ref{sec:conclusions} concludes this paper with some remarks and future work.

\section{Summarization\label{sec:related-work}}
Several algorithms for both generic and music summarization have been proposed.
However, music summarization algorithms were developed to extract an audible
summary so that any person can listen to it coherently. Our focus is on
automatic consumption, so coherence and clarity are not mandatory requirements
for our summaries.

LexRank \cite{Erkan2004} and TextRank \cite{Mihalcea2004} are centrality-based
methods that rely on the similarity between every sentence. These are based on
Google's PageRank \cite{Brin1998} algorithm for ranking web pages and are
successfully applied in text summarization. GRASSHOPPER \cite{Zhu2007} is
another method applied in text summarization, as well as social network
analysis, focusing on improving diversity in ranking sentences. \ac{MMR}
\cite{Zechner2000,Murray2005}, applied in speech summarization, is a
query-specific method that selects sentences according to their similarity to
the query and to the sentences previously selected. \ac{LSA} \cite{Gong2001} is
another method used in text summarization based on the mathematical technique
\ac{SVD}.

Music-specific summarization structurally segments songs and then selects which
segments to include in the summary. This segmentation aims to extract meaningful
segments (e.g. chorus, bridge). \cite{Chai} presents two approaches for
segmentation: using a \ac{HMM} to detect key changes between frames and
\ac{DTW} to detect repeating structure. In \cite{Cooper2003}, segmentation is achieved
by correlating a Gaussian-tempered ``checkerboard'' kernel along the main diagonal
of the similarity matrix of the song, outputting segment boundaries. Then, a
segment-indexed similarity matrix is built, containing the similarity between
every detected segment. \ac{SVD} is applied to that matrix to find its
rank-K approximation. Segments are, then, clustered to output the song's
structure. In \cite{Peeters2002,Peeters2003}, songs are segmented in 3
stages. First, a similarity matrix is built and it is analyzed for fast changes,
outputting segment boundaries. These segments are clustered to output the
``middle states''. Finally, an \ac{HMM} is applied to these states, producing
the final segmentation. These algorithms then follow some strategies to select
the appropriate segments. \cite{Chu2000} groups (based on the \ac{KL}
divergence) and labels similar segments of the song and then the summary is
generated by taking the longest sequence of segments belonging to the same
cluster. In \cite{Cooper2002,Glaczynski2011}, a method called Average Similarity
is used to extract a thumbnail $L$ seconds long that is most similar to the
whole piece. Another method for this task is the Maximum Filtered Correlation
\cite{Bartsch2005} which starts by building a similarity matrix and then a
filtered time-lag matrix, which has the similarity between extended segments
embedded in it. Finding the maximum value in the latter is finding the starting
position of the summary.

To apply generic summarization algorithms to music, first we need to segment the
song into musical words/terms. This fixed segmentation differs a lot from the
structural segmentation used in music-specific algorithms. Fixed segmentation
does not take into account the human perception of musical structure. It simply
allows us to look at the variability and repetition of the signal and use them
to find the most important parts. Structural segmentation aims to find
meaningful segments (to people) of the song so that we can later select those
segments to include in the summary. This type of segmentation often leads to
audible summaries which violate copyrights of the original songs. Fixed
segmentation combined with generic summarization algorithms may help evade those
issues.

In the following sections we review the algorithms we chose to evaluate:
Average Similarity, \ac{MMR}, LexRank, and \ac{LSA}.

\subsection{Average Similarity\label{sub:avg-sim}}
This approach to summarization has the purpose of finding a fixed-length
continuous music segment, of duration $L$, most similar to the entire song. This
method was introduced in \cite{Cooper2002} and later used in other research
efforts such as \cite{Glaczynski2011}.

The method consists of building a similarity matrix for the song and calculating
an aggregated measure of similarity between the whole song and every $L$ seconds
long segment.

In \cite{Cooper2002}, 45 \ac{MFCC}s are computed but only the 15 with highest
variance are kept. The cosine distance is used to calculate pairwise similarities.

In \cite{Glaczynski2011}, the first 13 \ac{MFCC}s and the spectral centre of
gravity (sound ``brightness'') are used. The Tchebychev distance was selected for
building the similarity matrix.

Once the similarity between every frame is calculated, we build a similarity
matrix $S$ and embed the similarity values between feature vectors $v_{i}$ and
$v_{j}$ in it: $S\left(i,j\right)=s\left(v_{i},v_{j}\right)$.

The average similarity measure can be calculated by summing up columns (or rows,
since the similarity matrix is symmetric) of the similarity matrix, according to
the desired summary length $L$, starting from different initial frames. The
maximum score will correspond to the segment that is most similar to the whole
song. To find the best summary of length $L$, we must compute the score
$Q_{L}\left(i\right)$:

\begin{equation}
Q_{L}\left(i\right)=\bar{S}\left(i,i+L\right)=\frac{1}{NL}\sum_{m=i}^{i+L}\sum_{n=1}^{N}S\left(m,n\right)
\end{equation}

$N$ is the number of frames in the entire piece. The index $1\leq i\leq
\left(N-L\right)$ of the best summary starting frame is the one that maximizes
$Q_{L}\left(i\right)$.

The evaluations of this method in the literature are subjective (human)
evaluations that take into account whether the generated summaries include the
most memorable part(s) of the song \cite{Cooper2002}. Other evaluations are
averages of scores given by test subjects, regarding specific qualities of the
summary such as Clarity, Conciseness and Coherence \cite{Glaczynski2011}.

\subsection{\acl{MMR}\label{sub:mmr}}
\ac{MMR}~\cite{Carbonell1998}, selects sentences from the signal
according to their relevance and to their diversity against the already selected
sentences in order to output low-redundancy summaries. This approach has been
used in speech summarization \cite{Zechner2000,Murray2005}. It is a
query-specific summarization method, though it is possible to produce generic
summaries by taking the centroid vector of all the sentences (as in
\cite{Murray2005}) as the query.

\ac{MMR} iteratively selects the sentence $S_i$ that maximizes the following
mathematical model:

\begin{equation}
\lambda\left({Sim_{1}}\left(S_{i},Q\right)\right)-\left(1-\lambda\right)\max_{S_{j}}Sim_{2}\left(S_{i},S_{j}\right)
\end{equation}

$Sim_{1}$ and $Sim_{2}$ are the, possibly different, similarity metrics;
$S_{i}$ are the unselected sentences and $S_{j}$ are the previously selected
ones; $Q$ is the query and $\lambda$ is a configurable parameter that allows the
selection of the next sentence to be based on its relevance, its diversity or a
linear combination of both. Usually sentences are represented as \ac{TF-IDF}
scores vectors. The cosine similarity is frequently used as $Sim_{1}$ and $Sim_{2}$.

\subsection{LexRank\label{sub:lexrank}}
LexRank \cite{Erkan2004} is a centrality-based method that relies on the
similarity for each sentence pair. This centrality-based method is based on
Google's PageRank \cite{Brin1998} algorithm for ranking web pages. The output is
a list of ranked sentences from which we can extract the most central
ones to produce a summary.

First, we compare all sentences, normally represented as \ac{TF-IDF} scores
vectors, to each other using a similarity measure. LexRank uses the cosine
similarity. After this step, we build a graph where each sentence is a vertex
and edges are created between every sentence according to their pairwise
similarity. Usually, the similarity score must be higher than some threshold to
create an edge. LexRank can be used with both weighted and unweighted edges.
Then, we perform the following calculation iteratively for each vertex until
convergence is achieved (when the error rate of two successive iterations is
below a certain threshold for every vertex):

\begin{equation}
S\left(V_{i}\right)=\frac{\left(1-d\right)}{N}+ S_{1}\left(V_i\right)
\end{equation}
\begin{equation}
S_{1}\left(V_{i}\right)=d\times\sum_{V_{j}\in
adj\left[V_{i}\right]}\frac{Sim\left(V_{i},V_{j}\right)}{\sum_{V_{k}\in adj\left[V_{j}\right]}Sim\left(V_{j},V_{k}\right)}S\left(V_{j}\right)
\end{equation}

$d$ is a damping factor to guarantee the convergence of the method, $N$ is
the total number of vertices and $S\left(V_{i}\right)$ is the score of vertex
$i$. This is the case where edges are weighted. When using unweighted edges, the
equation is simpler:

\begin{equation}
S\left(V_{i}\right)=\frac{\left(1-d\right)}{N}+d\times\sum_{V_{j}\in
adj\left[V_{i}\right]}\frac{S\left(V_{j}\right)}{D\left(V_{j}\right)}
\end{equation}

$D\left(V_{i}\right)$ is the degree (i.e., number of edges) of vertex $i$. We
can construct a summary by taking the highest ranked sentences until a certain
summary length is reached.

This method is based on the fact that sentences recommend each other. A
sentence very similar to many other sentences will get a high score.
Sentence score is also determined by the score of the sentences recommending it.

\subsection{\acl{LSA}\label{sub:lsa}}
\ac{LSA} is based on the mathematical technique \ac{SVD} that was first
used for text summarization in \cite{Gong2001}. \ac{SVD} is used to
reduce the dimensionality of an original matrix representation of the text. To perform
\ac{LSA}-based text summarization, we start by building a T terms by N sentences
matrix A.

Each element of A, $a_{ij}=L_{ij}G_{i}$, has two weight components: a local
weight and a global weight. The local weight is a function of the number of
times a term occurs in a specific sentence and the global weight is a function
of the number of sentences that contain a specific term.

Applying \ac{SVD} to matrix A will result in a decomposition formed by three
matrices: $U$, a $T\times N$ matrix of left singular vectors (its columns);
$\Sigma$, a $N\times N$ diagonal matrix of singular values; and $V^{T}$, a
$N\times N$ matrix of right singular vectors (its rows): $A=U\Sigma V^{T}$.

Singular values are sorted by descending order in matrix $\Sigma$ and are used
to determine topic relevance. Each latent dimension corresponds to a topic. We
calculate the Rank $K$ approximation by taking the first $K$ columns of $U$, the
$K\times K$ sub-matrix of $\Sigma$ and the first $K$ rows of $V^{T}$. We can
extract the most relevant sentences by iteratively selecting sentences
corresponding to the indices of the highest values for each (most relevant)
right singular vector.

In \cite{Steinberger2004}, two limitations of this approach are discussed:
the fact that $K$ is equal to the number of sentences in the summary, which, as
it increases, tends to include less significant sentences; and that sentences
with high values in several dimensions (topics), but never the highest, will
never be included in the summary. To compensate for these problems, a sentence
score was introduced and $K$ is chosen so that the $K^{th}$ singular value does
not fall under half of the highest singular value:
$score\left(j\right)=\sqrt{\sum_{i=1}^{k}v_{ij}^{2}\sigma_{i}^{2}}$.

\section{Experiments\label{sec:experiments}}
To evaluate these algorithms on music, we tested their impact on a Fado
classifier. This classifier simply classifies a song as Fado or non-Fado. Fado
is a Portuguese music genre whose instrumentation usually consists solely of
stringed instruments, such as the classical guitar and the Portuguese guitar.
The classifier is a \ac{SVM} \cite{Chang2011}.

The features used by the \ac{SVM} consist of a 32-dimensional vector per song,
which is a concatenation of 4 features: average vector of the first 13
\ac{MFCC}s of the song; \ac{RMS} energy; high frequencies 9-dimensional
rhythmic features; and low frequencies 9-dimensional rhythmic features.

These rhythmic features are computed based on the \ac{FFT} coefficients on the
20 Hz to 100 Hz range (low frequencies) and on the 8000 Hz to 11025 Hz range (high
frequencies). Assuming $v$ is a matrix of FFT coefficients with frequency
varying through columns and time through lines, each component of the
9-dimensional vector is: $maxamp$: max of the average $v$ along time; $minamp$:
min of the average $v$ along time; number of $v$ values above 80\% of
$maxamp$; number of $v$ values above 15\% of $maxamp$; number of $v$ values
above $maxamp$; number of $v$ values below $minamp$; mean distance between
peaks; standard deviation of distance between peaks; max distance between
peaks.

These features capture rhythmic information in both low and high frequencies.
Fado does not have much information in the low frequencies as it does not
contain, for example, drum kicks. However, due to the string instruments used,
Fado information content is higher in the high frequencies, making these
features good for distinguishing it from other genres.

We used 2 datasets in our experiments which consist of 500 songs from which
half of them are Fado songs and the other half are not. The 250 Fado songs are
the same in both datasets. The datasets are encoded in mono, 16-bit, 22050 Hz
Microsoft WAV files. We will make the post-summarization datasets available upon request.

We used 5-fold cross validation when calculating classification performance. The
classification performance was calculated first for the beginning, middle and
end sections (of 30s) of the songs to get a baseline and then we compared it
with the classification using the summaries (also 30s) for each parameter
combination and algorithm.

For feature extraction we used OpenSMILE's \cite{opensmile2013} implementation,
namely, to extract \ac{MFCC} feature vectors. We also used the Armadillo
library \cite{armadillo2010} for matrix operations and the Marsyas
library \cite{marsyas1999} for synthesizing the summaries.

For Average Similarity, we experimented with 3 different frame sizes (0.25, 0.5,
and 1 s) with both 50\% and no overlap. We also experimented with \ac{MFCC}
vector sizes of 12 and 24.

To use the generic summarization algorithms, however, we need additional
processing steps. We adapted those algorithms to the music domain by mapping the
audio signal frames (represented as \ac{MFCC} vectors) to a discrete representation
of words and sentences. For each piece being summarized, we cluster all of its
frames using the mlpack's \cite{mlpack2013} K-Means algorithm implementation
which calculates the vocabulary for that song (i.e., each frame is now a word
from that vocabulary). Then, we segment the whole piece into fixed-size sentences
(e.g., 5-word sentences). This allows us to represent each sentence as a vector
of word occurrences/frequencies (depending on the type of weighting chosen)
which lets us compare sentences with each other using the cosine distance.

In our implementation of \ac{MMR}, we calculate the similarity between every
sentence only once and then apply the algorithm until the desired summary length
is reached. We experimented using 3 different values for $\lambda$ (0.3, 0.5 and
0.7) and 4 different weighting types: raw (counting of the term), binary
(presence of the term), \ac{TF-IDF} and ``dampened'' \ac{TF-IDF} (same as
\ac{TF-IDF} but takes logarithm of TF instead of TF itself).

The damping factor used in LexRank was 0.85 and the convergence threshold
was set to 0.0001. We also calculated the similarity between every sentence only
once, applying the iterative algorithm and picking sentences until the desired
summary length is reached. We also tested LexRank using the same weighting types
as for \ac{MMR}.

We used Armadillo's \cite{armadillo2010} implementation of the \ac{SVD}
operation to implement \ac{LSA}. After sentence/word segmentation, we apply
\ac{SVD} to the term by sentences matrix (column-wise concatenation of all
sentence vectors). We then take the rank-K approximation of the decomposition
where the $K$th singular value is not smaller than half of the
$\left(K-1\right)$th singular value. Then, we calculate the sentence score (as
explained in section \ref{sub:lsa}) for each sentence and pick sentences
according to that ranking until the desired summary length is reached. We tested
\ac{LSA} with both raw and binary weighting.

We tested \ac{MMR}, LexRank, and \ac{LSA}, with all combinations of the
following parameter values: frame size of 0.5s with no overlap and with 50\%
(0.25s hops) overlap; vocabulary size of 25, 50, and 100 words; and sentence
size of 5, 10, and 20 words. We used \ac{MFCC} vectors (of size 12) as features for
these experiments, they are widely used in many MIR tasks including music
summarization in \cite{Cooper2003,Chu2000,Cooper2002,Glaczynski2011}.

\section{Results\label{sec:results}}
We present only the most interesting results, since we tried many different
parameter combinations for each algorithm. The Frame/Hop Size columns
indicate the frame/hop sizes in seconds, which can be interpreted as overlap
(e.g., the pair 0.5, 0.25 stands for frames of 0.5s duration with a hop size of
0.25s, which corresponds to a 50\% overlap between frames). The classification
accuracy results for the 30s contiguous segments which constitute the baseline
are 95.8\%, 96.2\%, and 94\% for the beginning, middle, and end sections,
respectively, on dataset 1 and 85.2\%, 92\%, and 90.4\%, on dataset 2.

The Average Similarity algorithm was successful in improving classification
performance on dataset 1 (98.8\% as maximum accuracy obtained with frame size of
0.5 s, no overlap, 24 \ac{MFCC}s), but not on dataset 2 (90.8\% maximum accuracy
with frame size of 0.25 s, no overlap, 12 \ac{MFCC}s).

In table \ref{tab:results}, we can see that although not all parameter
combinations for \ac{MMR} yielded an increase in classification performance on
both datasets, some combinations did do that. For example, the best combination on
the dataset 1 yielded 100\% accuracy but on dataset 2 it yielded only 90.8\%
which is lower than the baseline (92\%). However, all other parameter
combination presented in those tables yield a better result than the baseline
for both datasets. We also noticed that smaller values of $\lambda$ would
result in worse accuracy scores.

We can also see that the best parameter combination for LexRank on dataset 1
was also the best on dataset 2. Besides that, all other presented combinations
are better when compared to the corresponding baseline, which suggests that these
parameter combinations might also be good for other datasets.

Our experiments show that \ac{LSA} works best with binary weighting when applied
to music. This has to do with the fact that some musical sentences, namely, at
the beginning of the songs, are strings with very few repeating terms, which
increases term-frequency scores. Moreover, those terms might not even appear
anywhere else in the song which will, in turn, decrease the document frequency
of the term, thus increasing the inverse document frequency score. These issues
are detected when \ac{LSA} chooses those (unwanted) sentences because they will
have a high score on a certain latent topic. The binary weighting alleviates
these problems because we only check for the presence of a term (not its
frequency) and the document frequency of that term is not taken into account.
\ac{LSA} also achieved results above the baseline (table~\ref{tab:results}).

\begin{table}[htb]
\caption{MMR, LexRank and LSA (\#\ac{MFCC} = 12)}
\begin{center}
\begin{tabular}{c|c|c|c|c|c|c}
\begin{tabular}[c]{@{}c@{}}Frame\\ Size\end{tabular} &
\begin{tabular}[c]{@{}c@{}}Hop\\ Size\end{tabular} &
\begin{tabular}[c]{@{}c@{}}Vocab.\\ Size\end{tabular} &
\begin{tabular}[c]{@{}c@{}}Sentence\\ Size\end{tabular} & Weighting & $\lambda$
& Accuracy\\
\hline \multicolumn{7}{c}{MMR on Dataset 1}\\
\hline \begin{tabular}[c]{@{}c@{}}0.5\end{tabular} & 0.5 & 50 & 5 &
\begin{tabular}[c]{@{}c@{}}dampTF\end{tabular} & 0.7 & 100\%\\
\begin{tabular}[c]{@{}c@{}}0.5\end{tabular} & 0.25 & 100 & 5 &
\begin{tabular}[c]{@{}c@{}}Binary\end{tabular} & 0.7 & 99.2\%\\
\begin{tabular}[c]{@{}c@{}}0.5\end{tabular} & 0.5 & 25 & 5 & Binary & 0.5 &
97.2\%\\
\begin{tabular}[c]{@{}c@{}}0.5\end{tabular} & 0.5 & 25 & 10 & dampTF
& 0.7 & 97.6\%\\
\hline \multicolumn{7}{c}{MMR on Dataset 2}\\
\hline \begin{tabular}[c]{@{}c@{}}0.5\end{tabular} & 0.5 & 50 & 5 &
\begin{tabular}[c]{@{}c@{}}dampTF\end{tabular} & 0.7 & 90.8\%\\
\begin{tabular}[c]{@{}c@{}}0.5\end{tabular} & 0.25 & 100 & 5 &
\begin{tabular}[c]{@{}c@{}}Binary\end{tabular} & 0.7 & 93.4\%\\
\begin{tabular}[c]{@{}c@{}}0.5\end{tabular} & 0.5 & 25 & 5 & Binary & 0.5 &
93.4\%\\
\begin{tabular}[c]{@{}c@{}}0.5\end{tabular} & 0.5 & 25 & 10 & dampTF
& 0.7 & 93.4\%\\
\hline \multicolumn{7}{c}{LexRank on Dataset 1}\\
\hline \begin{tabular}[c]{@{}c@{}}0.5\end{tabular} & 0.5 & 25 & 5 & dampTF & - &
99\%\\
\begin{tabular}[c]{@{}c@{}}0.5\end{tabular} & 0.25 & 100 & 20 & Binary & - &
97.4\%\\
\begin{tabular}[c]{@{}c@{}}0.5\end{tabular} & 0.5 & 25 & 10 &
\begin{tabular}[c]{@{}c@{}}dampTF\end{tabular} & - & 97.6\%\\
\begin{tabular}[c]{@{}c@{}}0.5\end{tabular} & 0.5 & 25 & 10 & Raw & - & 97.6\%\\
\hline \multicolumn{7}{c}{LexRank on Dataset 2}\\
\hline \begin{tabular}[c]{@{}c@{}}0.5\end{tabular} & 0.5 & 25 & 5 & dampTF & - &
94\%\\
\begin{tabular}[c]{@{}c@{}}0.5\end{tabular} & 0.25 & 100 & 20 & Binary & - &
93.8\%\\
\begin{tabular}[c]{@{}c@{}}0.5\end{tabular} & 0.5 & 25 & 10 &
\begin{tabular}[c]{@{}c@{}}dampTF\end{tabular} & - & 93.8\%\\
\begin{tabular}[c]{@{}c@{}}0.5\end{tabular} & 0.5 & 25 & 10 & Raw & - & 93.4\%\\
\hline \multicolumn{7}{c}{LSA on Dataset 1}\\
\hline \begin{tabular}[c]{@{}c@{}}0.5\end{tabular} & 0.5 & 100 & 20 & Binary & -
& 99.6\%\\
\begin{tabular}[c]{@{}c@{}}0.5\end{tabular} & 0.5 & 25 & 10 & Binary & - &
99.4\%\\
\begin{tabular}[c]{@{}c@{}}0.5\end{tabular} & 0.5 & 50 & 10 & Binary & - &
96.6\%\\
\begin{tabular}[c]{@{}c@{}}0.5\end{tabular} & 0.25 & 25 & 20 & Binary & - &
97\%\\
\hline \multicolumn{7}{c}{LSA on Dataset 2}\\
\hline \begin{tabular}[c]{@{}c@{}}0.5\end{tabular} & 0.5 & 100 & 20 & Binary & -
& 91.2\%\\
\begin{tabular}[c]{@{}c@{}}0.5\end{tabular} & 0.5 & 25 & 10 & Binary & - &
93.4\%\\
\begin{tabular}[c]{@{}c@{}}0.5\end{tabular} & 0.5 & 50 & 10 & Binary & - &
93.4\%\\
\begin{tabular}[c]{@{}c@{}}0.5\end{tabular} & 0.25 & 25 & 20 & Binary & - &
92.8\%
\end{tabular}
\end{center}
\label{tab:results}
\end{table}

\section{Conclusions and Future Work\label{sec:conclusions}}
We evaluated summarization through classification for \ac{MMR}, LexRank, and
\ac{LSA} in the music domain. More experimenting should be done to find a set of
parameter combinations that will work for most music contexts. Future work
includes testing other summarization algorithms, other similarity metrics,
other types of features and other types of classifiers. The use of Gaussian
Mixture Models may also help in finding more ``natural" vocabularies and Beat
Detection might be used to find better values for fixed segmentation.

\cleardoublepage
\bibliographystyle{IEEEtran}
\IEEEtriggeratref{13}
\bibliography{on-the-application-of-generic-summarization-algorithms-to-music}

\end{document}